\documentclass{aastex}
\include{epsf}
\usepackage{emulateapj5}
\begin{document}
%%%%%%%%%%%%%%%%%%%%%%%%%%%%%%%%%%%%%%%%
% DEFINITIONS			       % 
%%%%%%%%%%%%%%%%%%%%%%%%%%%%%%%%%%%%%%%%
\def\SZ{Sunyaev-Zel'dovich}
\def\msun{M$_{\odot}$}
\def\ie{i.e.}
\def\eg{e.g.}
\def\etal{et al.}
\def\hou{km s$^{-1}$ Mpc $^{-1}$}
\def\omega0{$\Omega_\circ$}
\def\Ho{$H_{\circ}$}
\def\rc{$\theta_c$}
\def\be{$\beta$}
\def\dt{$\Delta T(0)$}
\def\dchi{$\Delta\chi^2$}
\def\da{$D_A$}
\def\rms{$r.m.s.$}
\def\uv{$u$-$v$}
\def\omegam{$\Omega_M$}
\def\omegal{$\Omega_\Lambda$}
%%%%%%%%%%%%%%%%%%%%%%%%%%%%%%%%%%%%%%%%
%                                      % 
%%%%%%%%
\submitted{to appear in Astrophysical Journal v538 Aug 1, 2000}
\title{The Sunyaev-Zel'dovich Effect in Abell 370}

\author{Laura Grego\altaffilmark{1}, John E. Carlstrom\altaffilmark{2}, 
Marshall K. Joy\altaffilmark{3}, Erik D. Reese\altaffilmark{2}, Gilbert P. 
Holder\altaffilmark{2}, Sandeep Patel\altaffilmark{4}, Asantha R. 
Cooray\altaffilmark{2}, and William L. Holzapfel\altaffilmark{5}}\hfill\break
\altaffiltext{1}{Harvard-Smithsonian Center for Astrophysics, 60 Garden
St., Cambridge, MA 02138}
\altaffiltext{2}{Department of Astronomy \& Astrophysics, 5640 S. Ellis Ave., 
University of Chicago, Chicago, IL 60637}
\altaffiltext{3}{Space Science Laboratory, ES84, NASA Marshall Space Flight 
Center, Huntsville, AL 35812}
\altaffiltext{4}{Department of Astronomy, University of Alabama, Huntsville, 
AL 35899}
\altaffiltext{5}{Department of Physics, University of California, Berkeley, CA 
94720}
\authoraddr{Harvard-Smithsonian Center for Astrophysics, 60 Garden
St., Cambridge, MA 02138}

\begin{abstract}
We present interferometric measurements of the Sunyaev-Zel'dovich (SZ) effect 
towards the galaxy cluster Abell 370.  These measurements, which directly 
probe the pressure of the cluster's gas, show the gas distribution to be strongly aspherical, as do the x-ray and gravitational lensing observations. 
We calculate the cluster's gas mass fraction in two ways.  We first compare the gas mass derived from the SZ measurements to the lensing-derived 
gravitational mass near the critical lensing radius.  We also calculate
the gas mass fraction from the SZ data by deprojecting the
three-dimensional gas density
distribution and deriving the total mass under the assumption that the
gas is in hydrostatic equilibrium (HSE).  
We test the assumptions in the HSE method by comparing the total cluster 
mass implied by the two methods and find that they agree within the
errors of the measurement.  We discuss the possible systematic errors
in the gas mass fraction measurement and the constraints it places on
the matter density parameter, \omegam.

\end{abstract}
\section{Introduction}
Clusters of galaxies, by virtue of being the largest known virialized objects, 
are important probes of large scale structure and can be used to test cosmological models.  
Rich clusters are extremely massive, $\sim$$\ 10^{15} M_{\odot}$, as indicated by the presence of strongly
 gravitationally lensed background galaxies and by the deep gravitational
potential necessary to explain both the large velocity dispersion ($>$ 1000 km s$^{-1}$) in the member galaxies
 and the high measured temperature ($> 5$ keV) of the ionized intracluster gas.  
Dynamical mechanisms for segregating baryonic matter from dark matter on these mass scales are difficult to reconcile with observations
and standard cosmological models, and so within the virial radius 
the mass composition of clusters is expected to reflect the universal mass 
composition.  
Under the fair sample hypothesis, a cluster's gas mass fraction, which is a lower limit to the its 
baryonic mass fraction, 
is then a lower limit to the universal baryon fraction, i.e., $f_{gas} \le f_B$.  

The luminous baryonic content of galaxy clusters is mainly contained 
in the gaseous intracluster medium (ICM).  The gas mass is 
nearly an order of magnitude 
larger than the mass in optically observed galaxies (\eg, White \etal\ 1993, Forman \& Jones 1982).  Hence, the gas mass is not only a lower 
limit to the cluster's baryonic mass, it is a reasonable estimate of it.  

The intracluster medium has largely been studied through observations of its x-ray emission.  The ICM is hot, with 
electron temperatures, $T_e$, from $\sim$5 to 15 keV; rarefied, with peak electron 
number densities of $n_e \simeq 10^{-3}$ ${\rm cm}^{-3}$; and cools
slowly ($t_{cool}> t_{Hubble}$), mainly via thermal Bremsstrahlung
in the x-ray band.
The x-ray surface brightness is proportional
to the emission measure, $S_x \propto \int n_e^2 \Lambda(T_e) dl$, 
where the integration is along the
line of sight, and so, under simplifying assumptions, the gas mass can be calculated from an x-ray image 
deprojection and the measured gas temperature.
Since the sound crossing time of the cluster gas is much less than the
dynamical time, one may reasonably assume that, in the absence of a
recent merger, the cluster gas is
relaxed in the cluster's potential.  The total binding mass can be
extracted from the gas density and temperature distribution under this
assumption. A significant body of work exists in which the gas
mass fraction, $f_g$, is measured in this way, with $f_g$ measurements out to radii of 1 Mpc or more (White
\& Fabian 1995; David \etal\ 1995; Neumann \& Bohringer, 1997; Squires
\etal\ 1997; Mohr, Mathiesen \& Evrard 1999).  The mean cluster gas mass fraction
within approximately the virial radius was calculated in Mohr \etal\
(1999) to be (0.0749 $\pm 0.0005) h^{-3/2}$.  Here, and throughout the
paper, we assume the value of the Hubble constant to be $H_\circ=100
h$ km s$^{-1}$ Mpc$^{-1}$.

To derive the gas mass fraction from x-ray imaging data, one is required to deproject the
surface brightness into a model for the density distribution. 
As the x-ray emission is proportional to the square of the gas density, the gas mass 
measurement can be biased by clumped, multi-phase gas, should it be present.
Also, the emission from the cores of relaxed clusters may be dominated 
by cooling flows, which complicate the interpretation of the x-ray data and
may bias the result strongly if not taken into account (Allen 1998; Mohr \etal\ 1999).
In addition, the x-ray surface brightness is diminished in proportion
to its distance; $S_x \propto 1/(1+z)^4$, where $z$ is the redshift of
the cluster, neglecting experiment-specific K-corrections, and so it becomes increasingly difficult to make sensitive 
x-ray measurements of the ICM as the cluster redshift increases.  
We present a scheme for measuring the gas mass fractions with the
Sunyaev-Zel'dovich effect which is different from the x-ray method in a
number of ways, and also provides an independent measurement of
$f_g$.

The Sunyaev-Zel'dovich (SZ) effect is a spectral distortion of Cosmic Microwave Background 
(CMB) radiation due to scattering of CMB photons
by hot plasma (Sunyaev \& Zel'dovich 1970). The SZ effect can be detected significantly in galaxy clusters, where the ionized intracluster gas serves as the scattering medium.  
A small fraction, $\leq 1\%$, of CMB photons are inverse-Compton scattered and, on average, gain energy.
At frequencies less than about 218 GHz, the intensity of the CMB radiation
is diminished as compared to the unscattered CMB, and the SZ effect
is manifested as a brightness temperature decrement towards the cluster.  This decrement, $\Delta T_{SZ}$, has a magnitude proportional
to the total number of scatterers, weighted by their associated temperature,
 ${\Delta T_{SZ} \over T_{CMB}} \propto \int n_e T_e dl$, where $n_e$ is the number density of electrons, $T_e$ is the electron temperature, $T_{CMB}$ is the temperature of the CMB, and the integration is again
along the line of sight.  Note that this is simply proportional to the
integrated electron pressure.  
Also, the magnitude of the SZ
decrement is independent of redshift, so as the long as the cluster is
resolved (the experiment's characteristic beamsize is not larger than
the angle subtended by the cluster), the SZ effect can be measured 
towards arbitrarily distant clusters. 

A cluster's gas mass is directly proportional to its integrated SZ effect
if the gas is isothermal.  So under the isothermality condition, an image
deprojection is not strictly required to obtain the gas mass.  The
cluster's gas mass fraction can be calculated by comparing the
integrated SZ decrement, in effect a surface gas mass, to the total
cluster mass in the same volume.  The total mass can be measured with strong or weak gravitational lensing, for example.
The SZ images may also be deprojected to infer the three-dimensional
gas mass and the HSE mass.  
  Since the SZ decrement is directly propotional to the electron density, the SZ image deprojection will not be affected strongly by clumped gas. Thus, the 
cluster's gas mass fraction can then be measured as a function of
cluster radius as well.

Recent cluster gas mass fraction measurements from SZ effect observations are
presented in Myers \etal\ (1997).  In this work, the integrated SZ effect is measured using a single radio dish operating at centimeter
wavelengths.  The integrated SZ effect is used to normalize a model
for the gas density from
published x-ray analyses, and this gas mass is compared to the published total masses to determine the gas mass fraction.  For three nearby clusters, A2142, A2256 and the Coma
cluster, Myers \etal\ find a gas mass fraction
of $(0.061\pm0.011) h^{-1}$ at radii of 1-1.5 $h^{-1}$ Mpc; for the
cluster Abell 478, they report a gas mass fraction of $(0.16\pm0.014)h^{-1}$.

In this work, we describe a method to calculate cluster gas mass fractions
from interferometric SZ observations. Here, the shape parameters are
derived directly from the SZ dataset rather than from an x-ray image.
We apply this method to the SZ effect measurements towards the cluster
Abell 370, which were made as part of an SZ survey of distant
clusters, the first results of which were 
reported in Carlstrom, Joy \& Grego (1996), and are further reported in 
Carlstrom \etal\ (1997).  We choose this primarily because it has been studied at optical and x-ray
wavelengths, and so allows a comparison of the SZ data with
other observations.  Upon detailed investigation, it is apparent that this cluster is one of the
most difficult in our sample to analyze, with significant ellipticity and
complicated optical and x-ray structure;  as such, it serves as a test
of the gas mass fraction analysis method, which we plan to use on a large sample of clusters.

With the interferometric SZ measurements, complemented by observations at other wavelengths, 
we measure the cluster's gas mass fraction in two ways. First, we calculate the surface gas mass from the SZ measurements and 
 the surface total mass from strong gravitational lensing observations and
models.   Second, we measure the gas mass 
fraction in a manner similar to x-ray analyses.  The gas mass is
inferred from a deprojected model; the total mass is determined from
the spatial distribution of the gas under the isothermal hydrostatic
equilibrium (HSE) assumption.  
We test the assumptions made in the HSE analysis by comparing
the total cluster masses derived in the two methods.

The optical and x-ray observations of this cluster are discussed in Section~\ref{sec:optxray}, and the SZ observations are discussed in Section~\ref{sec:szobs}.  
The method for modeling the SZ data is presented in
Section~\ref{sec:model}, and the gas mass fraction results and the systematic uncertainties are discussed in Section~\ref{sec:analysis}.  The cosmological implications of the results and plans for future work are discussed in Section~\ref{sec:discussion}.

\section{Optical and X-ray Observations}
\label{sec:optxray}
The spatial distribution and redshifts of a number of Abell 370's constituent galaxies have been measured optically.  
Mellier \etal\ (1988), using their 29 best galaxy spectra, find Abell 370 has
 a mean redshift z = 0.374 and velocity dispersion $\sigma_v=1340^{+230}_{-150}\ {\rm km\ s^{-1}}$.  
These observations also show that the cluster is dominated by two
giant elliptical galaxies, at ($\alpha_{J2000} = 02^h\ 39^m\ 52.7^s$,
$\delta_{J2000} = -01^{\circ}\ 34'\ 20.3''$) and ($\alpha_{J2000} = 02^h\ 39^m\ 53.2^s$,
$\delta_{J2000} = -01^{\circ}\ 34'\ 57.6''$), with a projected separation of about $40
''$ in the north-south direction.  The major axes 
of these two dominant galaxies are also oriented nearly north--south.  The Mellier \etal\ optical isopleth map shows two density peaks, 
with similar north-south orientation, although these are not centered on the two elliptical galaxies.  
The 29 galaxy velocities do not appear to be distributed in a Gaussian
manner, although the statistics on this modest sample are not
conclusive.  Should this be significant, its
may suggest that the cluster is bimodal or that it is not yet fully
relaxed, and the assumption of a simple one-component model in
hydrostatic equilibrium would not be appropriate.
The mean velocities in the two isopleth peaks are not found to 
differ significantly, however, nor are the calculated barycenters of the two
peaks, and so it is difficult to make a firm conclusion on the state
of the cluster based on these data alone.
      
Abell 370 was the first cluster found to have an associated strongly gravitationally 
lensed arc $(z_{arc} = 0.725$; Soucail \etal\ 1988).
In addition to this giant arc, which apparently consists of multiple images of a single source, three faint pairs of images, each pair presumed to be lensed images of a single galaxy 
(Kneib \etal\ 1993), and a radial arc (Smail \etal\ 1996)
have been found to be associated with this cluster.  Gravitational lens models have been used to map 
the total mass distribution in the core region of Abell 370.  Kneib \etal\ (1993) find the 
position and elongation of the giant arc and the positions of the 
multiple images are best reproduced by a bimodal distribution of 
matter, with the two centers of mass nearly coincident
with the cluster's two large elliptical galaxies.  A subsequently 
discovered radial arc is consistent with this model without modifications
(Smail \etal\ 1996).  
%In a forthcoming paper, (Bezecourt \etal\ 1998), the 
%lensing model for this cluster will be further refined.

Abell 370 was observed with the High Resolution Imager (HRI) of the {\em ROSAT} x-ray satellite.
In the {\em ROSAT} bandpass (0.1-2.4 keV), x-ray Bremsstrahlung emission from a plasma at 
 gas temperatures typical of clusters depends only weakly on the gas 
temperature.  In Figure 1, we show an HRI image of A370, the 
result of $\sim$ 30 kiloseconds of observation time.
The bright source to the north, at $\alpha_{J2000}= 02^h\ 39^m\ 55^s$,  $\delta_{J2000}=-01^\circ\ 32'\ 33''$, is coincident with what appears 
in the Digital Sky Survey to be 
a nearby elliptical galaxy, NPM1G -01.0096 (Klemola, Jones \& Hanson 1987), and does not appear to be associated with
A370.  
The image indicates that the intra-cluster medium also is elongated in the north-south direction, and the emission 
is strongly peaked at the position of the southern dominant elliptical
galaxy, numbered 35 in Mellier \etal\ (1988), and is offset from the
center of the large scale emission.  
The cluster gas distribution does not appear clearly bimodal at this
resolution.

We determine the x-ray surface brightness from the ROSAT HRI image, 
which has been filtered to include only PHA channels 1-7 in order to 
reduce the background level (David \etal\ 1997) and blocked into 8$''
\times$ 8$''$ pixels.  A circular region with a radius of 24$''$
centered 
on the pointlike x-ray source to the north of the cluster was excluded
from our modeling.  
We fit a \be-model to the data; this model described in more detail in Section~\ref{sec:model},
which parametrizes the x-ray surface brightness with a core radius, \rc, and
a power law index, \be.  The surface brightness of a spheroidally
symmetric, isothermal gas distribution will have elliptical contours,
and so we also fit for the axis ratio.

We find the
best-fit axis ratio of the x-ray image to be 0.73$\pm0.03$, with major axis
exactly north-south, and the center position to be $\alpha_{J2000} =
02^h\ 39^m\ 53^s$, $\delta_{J2000}=-01^\circ\ 34'\ 33''$.  The best
fitting shape parameters for the elliptical model are \be= 0.70$\pm 0.03$ and
\rc = 85.0$\pm 3.5''$.  If the axis ratio is set to equal 1.0, the best fitting
shape parameters are \be = 0.72$\pm 0.03$ , \rc = $(70.1\pm 3.2)'' $ (90\%
confidence on a single parameter).

Abell 370 was also observed with the {\it Einstein} IPC for 4000
seconds.  In the IPC image, the northern point source is not
apparent.  This may be because the FWHM of the instrument point
spread function is about 90$''$ at the middle of the 0.4-4.0 keV
spectral range of the instrument. (It improves to about 60$''$ at
higher energies.)  The point source is offset by less than 120$''$ from
the center of the extended emission, and so may not be resolved in
this image.  It may also be explained by source variation; if the point source is an AGN, its surface brightness may have
significantly varied between the 1979 IPC observation and the
1994-1995 HRI observations.  The image looks less elliptical than does
the HRI image, but because the point source emission cannot be
excluded from the cluster emission in the IPC data, it is not useful 
for constraining the spatial distribution of the cluster gas.

%\placefigure{fig1}

Abell 370 has also been observed with the ASCA x-ray satellite.  
Using 37.5 kiloseconds of {\em ASCA} GIS/SIS observations, Mushotzky \& Scharf 
(1997) determined the emission-weighted average temperature of the cluster 
to be $7.13 ^{+1.05}_{-0.83}$ keV.   Analyzing the same data, 
Ota, Mitsuda, \& Fukazawa (1998) find the emission-weighted temperature to be $6.6^{+1.1}_{-0.9}$ keV. 
Both of these values are systematically lower than the value obtained
in the first analysis of these data by Bautz \etal\
(1994), who obtain a value of $8.8 \pm 0.8$ keV.  The discrepancy is 
presumably because the Mushotzky \etal\ and Ota \etal\ analyses used the 
values of the response matrices of the {\em ASCA} telescope and x-ray detectors 
which have been refined since the Bautz \etal\ analysis (M. Bautz, private communication).  For our analysis, 
we adopt the Ota \etal\ value of $6.6^{+1.1}_{-0.9}$ keV, because this published analysis includes
detailed discussion of the spectral fitting procedure.

The bright point source evident in the {\em ROSAT} HRI image has not been removed from the {\em ASCA} spectra
in any of the three published analyses.  If this emission originates in 
Bremsstrahlung emission from hot gas in the elliptical galaxy, its characteristic 
temperature is expected to be less than 1 or 2 keV (Sarazin \& White 1988 and references therein).  Since the
sensitivity of {\em ASCA} is optimized for higher energies than this, we would not expect the measured
temperature to be greatly contaminated.  We assess whether the source is extended, and find that it
is pointlike at the resolution of the HRI with a 2.4 $\sigma$
certainty. As the source is
unresolved,  this emission is also consistent with a galactic
cooling flow; if this were the case, the temperature again shouldn't be
greatly contaminated.  Should the emission originate from an AGN in the galaxy, though, the effect
on the measured temperature is more difficult to predict, and will depend strongly on the index of the power law of the AGN spectrum.  
 
Without further measurements, we cannot rule out that
the cluster's spectrum is contaminated by emission from an AGN.  It is also possible that
if the cluster is in
fact bimodal with two distinct subclusters, the measured emission-weighted
average $T_e$
will be a value between the temperatures of the two subclusters.
Lacking more definitive measurements, we continue with our assumption
that the gas is isothermal at $kT=6.6^{+1.1}_{-0.9}$ keV.  The estimated effects of the described uncertainties on the results are discussed in Section~\ref{subsec:systematics}.

We compare A370's measured line-of-sight velocity dispersion with our
adopted gas temperature, and find $\beta_{spec} \equiv {\mu m_p\sigma^2_v}/{{k T_e}} = 1.87^{+1.02}_{-0.59}$.  If the energy per unit mass
contained in the galaxies is equal to that in the gas, and the cluster is spherical,
 we expect $\beta_{spec}$ to
be $\sim 1$.  If the temperature measurement is biased low, \eg, from an AGN, 
this will contribute to the $\beta_{spec}$ discrepancy.
Elongation of the cluster potential along the line of sight 
or a alignment of two (or more) subclusters will also enhance the 
measured $\beta_{spec}$.

The optical and x-ray observations suggest that A370 is not a relaxed
cluster, and may be heavily substructured, but due to insufficient signal-to-noise ratios and contaminating sources, these
observations still leave some ambiguity.

\section{SZ Observations}
\label{sec:szobs}
	We observed Abell 370 for 50 hours over nine days with the Owens Valley Radio Observatory (OVRO) Millimeter Array in 1996 August and 43 hours
over six days with the Berkeley-Illinois-Maryland Association (BIMA) Millimeter Observatory in 1997 
June-August.  We set our pointing center to $\alpha_{J2000}= 02^h\
39^m\ 52.5^s$, $\delta_{J2000}=$ $-01^{\circ}\ 34'\ 20''$, the position of the northern dominant elliptical galaxy.

	We outfitted the millimeter telescopes with centimeter-wavelength 
receivers to broaden the instrument's angular resolution to the 
large angular scales typical of clusters.  The receivers are based on low noise 
High Electron Mobility Transistor (HEMT) amplifiers (see Pospieszalski 1995), operating from 26 to 36 GHz.  The single-mode receivers 
are constructed to respond only to circularly polarized light.  
The receivers interface with each array's delay lines and correlator.  
The standard observational scheme entailed interleaving 20 minute cluster 
observations with observations of a strong reference source near the cluster, which allows monitoring of the instrumental amplitude and phase gain.  
The amplitude gain drifts were minimal; the gain changes were less than
a percent over a many hour track, and the average gains were quite stable from day to day.  The amplitude gains were 
calibrated by comparing the measured flux of Mars with that predicted by the Rudy model for 
Mars' whole disk brightness temperature (Rudy, 1987) at the observed frequency and the apparent
size of Mars as indicated by the Astronomical Almanac.
The Rudy model is a radiative transfer model with an estimated accuracy of 
{4\%} (90\% confidence) at centimeter wavelengths.   
Data taken when the projected baseline was within 3\% of the shadowing limit 
were not used.

%\placefigure{fig2}
   
	The OVRO Millimeter array consists of six 10.4 meter telescopes.  
The OVRO continuum correlator was used to correlate two 1 GHz bands centered at 
28.5 GHz and 30 GHz.  The system temperatures ranged from 40-60 K,
depending on elevation and atmospheric water content.  The primary
beams of the telescopes were measured holographically and can be
approximated as Gaussian with a full width at half maximum (FWHM) of 252$''$.  
The data were calibrated and edited using the MMA data reduction package 
(Scoville \etal\ 1993) taking care to remove data taken during poor weather or which show any anomalous phase jumps.

	We image the data using the DIFMAP package (Shepard, Pearson,
\& Taylor 1994).  Examining images of A370 made with projected
baselines greater than 2.0 $k{\lambda}$, we find a point source 45$''$ to the east of the map center (map center is the pointing center) in both continuum bands.  
Its measured flux density in the 28.5 GHz band, attenuated by the instrumental primary beam pattern, is +0.69 $\pm$ 0.10 mJy and is +0.84 $\pm$ 0.10 mJy in the 30 GHz band.  
This source, 0237-0147, is discussed further in Cooray \etal\ 1998.  
We combine the two continuum
channels, and find the best fitting model is a point source with flux density of $0.725 \pm 0.07$ mJy.  The point source model is removed from
the \uv\, or spatial frequency, data set in order to construct the SZ
decrement map. A Gaussian taper with FWHM of 1.2 $k\lambda$ is applied in the \uv\ plane to the
combined dataset and map was CLEANed, restricting the CLEAN components
to the central 200$''$, about the size of the decrement in the
primary.  The map presented in Figure 2a is made with
a restoring beam of Gaussian FWHM $59''$ x $86''$, with contour
intervals of 1.5 $\sigma$.  The $rms$ noise in the map is 50
$\mu$ Jy beam$^{-1}$, or 14.9 $\mu$K, and the integrated flux density of the source is $-1.54\pm0.17$ mJy.

We employed nine of the 6.1 meter telescopes of the BIMA array for SZ 
observations, using the same centimeter-wave receivers we used at OVRO.    
We center the 800 MHz bandwidth of the BIMA digital correlator at 28.5 GHz.  
We achieved system temperatures of 30-55 K, scaled to above the 
atmosphere, depending on elevation and atmospheric water content.  Holographic antenna pattern measurements were also made with this system. The 
primary beam at BIMA is nearly Gaussian with a $396''$ FWHM.  The data were edited and reduced using the 
MIRIAD data reduction program (Sault, Teuben \& Wright 1995), taking care to the 
remove any spurious interference from the spectral channels and dropping the 
low signal-to-noise channels at the spectral filter edges. 
%The fits are nearly Gaussian was quite good, with an r.m.s. residual of 
%0.011 Jy.  The 
%The $360''$ grid with $90''$ spacing was best fit by a Gaussian with major axis FWHM $382''$ and
%minor axis FWHM $379''$, also with r.m.s. residual of 0.011 Jy.
%We adopt a circularly symmetric Gaussian with FWHM = $6.3'$ when modeling the data.  

 Again using DIFMAP, we find the same point source at about $45''$ to the 
east of the pointing center in a high resolution (projected baselines 
longer than $1.4~ k\lambda$) image.  The point source measured at BIMA
in 1997 has flux density of +0.70 $\pm$ 0.17 mJy. The primary beam attenuation
 at the point source position in the BIMA system is $\sim6 \%$ less than 
the attenuation at this position at OVRO, and so the point source
observations are consistent with the flux being constant in time.  We model 
and subtract the point source from the data set, apply a Gaussian taper with FWHM of 0.8 $k\lambda$ in the \uv\ plane and construct a map of the decrement. The map was CLEANed, restricting the CLEAN components to the central 
$\sim 300''$ of the image, about the size of the primary beam.
The restoring beam used to make the map in Figure 2b. is a Gaussian
with FWHM $91''$ x $95''$, and the contours presented are at intervals
of 1.5 $\sigma$.  The $rms$ noise in
the map is 180 $\mu$Jy beam$^{-1}$, or 31.4 $\mu$K, and the integrated flux density of the
source is $-4.34\pm0.52$ mJy.

It is instructive to compare the general characteristics of the SZ images to the x-ray map.
At the resolution of both the OVRO and BIMA instruments, 
the gas is extended in the north-south 
direction, though
less markedly than in the x-ray map.  The substructure suggested
by the lensing and velocity dispersion data is not evident here. We 
produce maps from each of the datasets, increasing the image
resolution, but the gas still does not 
show significant substructure or bimodality. 
The SZ decrement distribution is not peaked at the pointing center, but
south of it, at a position about halfway between the two elliptical galaxies (this will be
quantified in the next section).  
The detailed appearance of the SZ map, especially the shapes of the least 
significant contours, depends on the specific method of point source removal 
and the CLEANing of the data, and so it is not useful to compare 
the spatially filtered and CLEANed SZ and x-ray images at such a level
of detail.  As we will discuss in the following section, the quantitative analysis of the SZ data is done using the \uv\ data directly.

\section{Modeling}
\label{sec:model}

 In order to assess the data more quantitatively, we fit a parametrized 
model to the SZ brightness distribution.  
Since the SZ data are taken in the spatial frequency, or \uv, domain, we do the model 
comparisons in \uv\ domain as well.

We measure the SZ temperature decrement in units of antenna temperature,
$T_a$, which relates an observed intensity change to a Rayleigh-Jeans
temperature change, $\Delta I_{\nu} = {2k\nu^2 \over c^2}\Delta T_a$.
In order to recover the true blackbody temperature decrement, the
$\Delta T_{RJ}$ measured and reported here should be multiplied by a
factor of 1.021, as we are not strictly in the Rayleigh-Jeans limit.
The SZ temperature decrement, ${\Delta T_{SZ}}\over{T_{CMB}}$, is proportional to the Compton $y$-parameter, 
\begin{equation} y = {{k \sigma_T}\over {m_e c^2}} \int {n_e(l) T_e(l) dl}, \label{eq:compton}\end{equation}
where $k$ is Boltzmann's constant,  $\sigma_T$ is the Thomson scattering cross section, $m_e$ is the electron mass, 
$n_e$ is the electron density, $T_e$ is the electron gas temperature, 
and the integral extends along
the line of sight ($dl$).
The proportionality depends on the observing frequency; it depends 
also on the electron
temperature when relativistic corrections are included\footnote{The
change in spectral intensity due to the Sunyaev-Zel'dovich effect is
calculated in Equation (4-8) Challinor \& Lasenby (1998): ${\Delta
\left(T_{SZ}/T\right)_{RJ} = {y x^2 e^x \over (e^x -1)^2}[ x \coth(x/2) - 4 + \theta_e f(x)],}$ where $x = {h\nu \over kT_e}$ and $\theta_e = {kT_e \over m_e c^2}$. The last term corrects for relativistic effects.  At 28.5 GHz, $f(x) = 3.58$.} (Rephaeli 1995, Challinor \& Lasenby 1998).

At 28.5 GHz, ${\Delta T_{RJ}\over T_{CMB}} = -1.92$ $y$ in the non-relativistic Rayleigh-Jeans approximation, where we adopt the value of the CMB of Fixsen \etal\ (1996) derived from the COBE FIRAS measurements, $T_{CMB}=2.728$ K.  Including the relativistic corrections for 
$kT_e = 6.6$ keV, ${\Delta T_{SZ}\over T_{CMB}} = -1.86\ y$.

We fit a simple model to the SZ data.  The \be-model (Cavaliere \& Fusco-Femiano 1976, 1978) is frequently used to fit the density profiles of galaxy clusters.  The spherically symmetric, isothermal \be-model describes $n_e$, the density distribution of the gas, as varying as a function of cluster radius, $r$:
\begin{equation}n_e(r) = n_{e\circ}\left(1 + {r^2 \over r_c^2}\right)^{-3\beta /2},\label{eq:density}\end{equation}
where $n_{e\circ}$ is the number density of the gas at the
center of the spheroidally symmetric gas distribution; $r_c$, the core
radius, is a characteristic size of the cluster;  and $\beta$ is the power law index.

Since the cluster appears elliptical in projection, we generalize the spherically-symmetric \be-model and allow the gas density profile to be spheroidally symmetric, \ie, with biaxial symmetry.  In the spheroidally symmetric model, the electron
number density in a prolate spheroid is a function of $\eta^2 = r^2 +
a^2 z^2$, where  $r$ and $z$ are the radial and height coordinates in
the cylindrical coordinate system, and $a < 1$ is the axis ratio of the two unique axes.  
If the spheroid is oblate, then the density is a function of $\eta^2 = a^2r^2 + z^2$, with $a < 1$.
The electron density then follows the distribution:
\begin{equation} n_e(\eta) = n_{e\circ} \left ( 1 +  {\eta^2 \over r_c^2} \right )^{-3\beta/2},\end{equation}
If the cluster gas is isothermal and its symmetry axis is in the plane of the sky, this electron density distribution leads to the following two-dimensional SZ temperature decrement:
\begin{equation} {\Delta T}\left ( \theta \right ) = \Delta T(0) \left( 1 +  {\theta_{\eta}^2 \over \theta_c^2} \right)^{{1\over 2} -{3\beta \over 2}},\label{eq:DT/T}\end{equation}
where $\theta_{\eta} = \eta/D_A $, $D_A$ is the angular diameter distance,
$\theta_c = r_c / D_A$, and $\Delta T(0)$ is the temperature decrement at zero projected radius,
\begin{equation}{\Delta T(0) \propto T_e n_{e\circ}\int \left( 1 + \left( {l \over r_c }\right)^2 \right)^{-{3\over 2}\beta}dl},\label{eq:DT0}\end{equation}
where the integral, $dl$ is along the line of sight.  Formally, this integral extends from the observer along the line of sight through the cluster infinitely; in practice, a cutoff radius for the cluster is used.  The \be-model distribution of the electron density projects to a \be-model distribution of the
SZ decrement if the system obeys spherical or ellipsoidal symmetry.

We perform a $\chi ^2$ analysis of the \be-model, by comparing
$\beta$-models to the combined BIMA and OVRO datasets.  We vary the
following parameters: centroid position, \be, \rc, axis ratio (defined to be $<$ 1), position angle (defined counter clockwise from north), and \dt.  The position
and flux density of the radio-bright point source are also fit.
The fitting procedure is conducted in several steps.  The model corresponding to each set of the seven cluster fit parameters and the three point source parameters is multiplied
 by the primary beam response.  The Fourier transform of the result is compared directly with the interferometer data.  We use the holographically determined primary beams 
when modeling the data, and the entire datasets are 
used to do the analysis.  The inner \uv\ radius cutoff is determined by the
shadowing limit, the limit where one telescope would partially block another, \ie, when the projected baseline is less than the diameter of a telescope dish.
  For the BIMA data this limit is $0.58\ k\lambda$ and for the OVRO
data it is $1\ k\lambda$.  The $\chi^2$ statistic is minimized, and
the best fit values are determined using a downhill simplex method.

Performing the fitting procedure, we find the best fit parameter values for A370 are: \rc\ = $93''$, \be\ = 1.77, 
\dt\ $= -609$ $\mu K$, and axis ratio = 0.64 with major axis exactly
north-south.  The best fit central position is $20''$ to the south of
the pointing center, at $\alpha_{J2000} = 02^h\ 39^m\ 53.2^s$,
$\delta_{J2000} = -01^{\circ}\ 34'\ 40.4''$, about halfway between
the two giant elliptical galaxies, and 7$''$ north of the fitted
centroid for the x-ray image.  The $\chi^2$ statistic for the best
fit values is 135798 for 136136 degrees of freedom, yielding a reduced
$\chi^2$ of 0.9975.  As points of comparison, we fit to the data a null model for the
cluster (with the point source component), which gives a reduced
$\chi^2$ of 0.9986; we also fit the
cluster to a negative point source, finding a reduced $\chi^2$ of 0.9986.

In order to determine the uncertainties in these fitted parameters, the
$\chi^2$ statistic is derived for a large range of \be, \rc, and \dt, 
keeping centroid position, position angle, and point source 
position and flux fixed to the best fit values.  Presented in Figure 3a is a contour plot of the \be\ and \rc\ fit results, where \dt\ is 
allowed to assume its best fit value at every pair of \be\ and \rc.
The axis ratio is left to assume its best fit value at each set of 
\be, \rc, and \dt; it
varies from 0.60 to 0.68 for points with \dchi\ $<$ 5 from the best fit point.  The centroid position, when left to assume
its best
fit value at each point, does not appreciably change, varying less
than 5$''$.  
%The gas fraction calculation results in Section~\ref{sec:analysis} do
%not differ if the central position and axis ratio are allowed to vary.
Figure 3b shows the contours derived when fitting the data to a
spherical model for the gas distribution, \ie, the axis ratio is fixed
to a value of 1.0;
in this case, the best fit parameter values are \rc\ = $34''$, \be\ = 0.86, 
\dt\ $= -785$ $\mu K$, the reduced $\chi^2$ statistic was 0.9976. 

\ The full line contours are marked for \dchi~ = 2.3, 4.61, and 6.17 which
indicate 68.3\%, 90.0\%, and 95.4\% confidence, respectively, for the
two-parameter fit.  The dashed lines indicate \dchi~$=1.0$, 2.71, and
6.63;  the projection onto the \be\ or \rc\ axis of the interval
contained by these contours indicate the 68.3\%, 90\% and 99\%
confidence interval on the single parameter.  The values of \dt\ for
which \dchi\ $<$ 1 at each (\be, \rc) point range about 15\% from
the best fit value.  This illustrates that \be\ and \rc\ are correlated strongly and are not individually well
constrained by these data.  This poor constraint is at least partly
due to the low declination of A370, which makes sampling non-redundant
spatial frequencies difficult at OVRO and BIMA. The shape parameters fit from
the x-ray data are not consistent with those from the SZ data.  This
could be a result of shocks and complexity in the gas phase which
affect the x-ray emission and SZ effect differently.  This discrepancy
is difficult to resolve with the information at hand; for the purpose
of the next section, investigating the gas mass fraction of the
cluster using the SZ effect, we use the SZ fitted parameters only.

%\placefigure{fig3}

We also fit the data with a \be-model which includes a truncation of the 
gas distribution at a given radius.  
With truncation radii from 300$''$ to 1000$''$, the shape parameters
(\be, \rc, \dt) best fit values do not change appreciably, much less
than the variation contained within the \dchi\ $<$ 1 region. The $\chi^2$ statistic changes only minimally for different cutoff radii, with \dchi\ less than 0.1.
This indicates
that the possible systematic uncertainty in the fitted parameters due
to modeling the data without a cutoff is not significant.

Since the optical and x-ray data suggest that Abell 370 may have a bimodal 
gravitational potential, 
we also fit the data with a pair of circular \be-models, each allowed
independent shape parameters and position.  The
best fit two-component model has one component centered 
at $\sim 50''$ south of the pointing center, 10$''$ south of the large 
southern elliptical galaxy; the best fit position of the second component is $\sim 5''$ north
of the pointing center, \ie, $5''$ north of the large northern
elliptical.
The southern component has best-fit parameters $\beta$ = 0.65,
\rc = 10$''$, and \dt\ = $-765$ $\mu$K; the northern component has
best-fit parameters \be\ = 0.83, \rc\ = 27$''$, and \dt\ = $-605$ $\mu$K.  
Adding a second component reduces the $\chi^2$ statistic by 2 when 3 new 
parameters are introduced (compared to the single $\beta$-model with
axis ratio and position angle free), and therefore
is not a significantly better fit to the data.  Allowing the axis
ratios of the two components to vary does not improve the fit.

\section{Analysis}
\label{sec:analysis}
\subsection{Comparison of SZ Gas Mass to the Strong Lensing Mass Estimate}
As indicated by Equation~\ref{eq:compton}, the SZ brightness at any point in the two-dimensional 
projected image is simply proportional to the integrated electron density 
along the line of sight, if the gas is isothermal.
Under the isothermal assumption, we can directly measure the total number of 
electrons in the gas contained in the
cylindrical volume of a chosen radius, the long axis of which is defined by the line of sight.  The total mass in the ionized phase can be calculated from this assuming a value for the number of nucleons per electron.  If the gas has solar metallicity, as measured by 
Anders \& Grevesse (1989), the nucleon/electron ratio is 1.16.  The nucleon/electron ratio changes less than 1\% for values of the metallicity from 0.1 to 1.0.

The interferometric measurements 
recover much of the total SZ decrement on the angular scales measured, \ie\, the
integrated flux from the data comprises 40-50\% (depending on the
instrument) of the integrated flux
in a best fit model to these data. And so the unmodeled interferometer data
provides a strong lower limit to the integrated SZ decrement at the
angular scales of interest.
The model is fitted to the \uv\ data in
order to estimate the full, two-dimensional decrement, and thence the
surface gas mass. 
This method does not assume anything about the state of the ICM other
than that it is isothermal.

We fit the SZ data in the three-dimensional parameter space {0.4 $< \beta <$ 4.0, 10$'' < r_c <
$250$''$, $-2000$ $\mu$K $< \Delta $T(0) $<$ $-200$ $\mu$K}, both for
elliptical and circular models.  At each
{\be, \rc, and \dt} point, we calculate the surface gas mass and the
$\chi^2$ statistic, from which confidence limits for the
surface gas mass are determined.  Although the 68\% confidence region
contains a large range of \be\ and \rc\ values within 68\% confidence, as was evident in Figure 3, 
the gas mass, which depends on all three gridded parameters, is constrained relatively well.  In Figure 4, we show
the derived surface gas mass at radius 65$''$ as it varies with \be\ and
\rc\ in the circular \be-model fitting.  The gas mass in this geometry
is $1.5^{+0.7}_{-0.6} \times 10^{13} h^{-2} M_{\odot}$. The temperature decrement is
allowed to assume its best fit value at each point for this figure,
although for the quantitative analysis, the full range of \dt\ is used.  The isomass surfaces, shown in greyscale,  
follow the shape of the confidence interval contours, indicating that sets
of parameters which fit the data well will predict the same gas mass.  This is
to be expected, since under the isothermal assumption, the SZ flux is directly
proportional to the gas mass, and the fit parameters must reproduce
the same observed flux at the angular scales where the flux is best
measured. 

We derive a surface gas mass to compare with the lensing mass using our fitted model parameters.  The lensed arc in Abell 370 has a radius of curvature of about 30$''$
centered nearly halfway between the dominant galaxies;  this center is at about the same position of the gas density centroid in the one component $\beta$ model.  
We calculate the surface gas mass in a cylindrical volume centered at
this position and with an elliptical (axis ratio = 0.64)
cross-section, using the fits to the elliptical \be-model.  We choose a major radius of 40$''$ since the lens model should be
most accurate near the lensed arc's radius. 
The SZ-derived surface gas mass with the elliptical cross-section is
$5.4^{+1.2}_{-1.0} \times 10^{12} h^{-2} M_{\odot}$, at 68\%
confidence.  Included in the error estimates are the uncertainty due
to the fit parameters, the SZ absolute calibration uncertainty, and
the gas temperature measurement uncertainty.  The uncertainty is not
appreciably smaller if we restrict the shape parameters to those which
typically found in x-ray image analyses of clusters, \ie, \be\ $<
1.5$.  The uncertainties of the masses derived from the elliptical
fits are smaller than those from the spherical fits because the
confidence contours follow the isomass contours more closely.

%\placefigure{fig4}

We also derive the gas mass in the model with two components.  The
best fitting two-component \be-model parameters are integrated in a
cylinder with radius 40$''$, centered on the centroid of the best fit single
component model.  This yields a gas mass of $4.84 \times 10^{12} h^{-2}
M_{\odot}$, consistent with the single component model.
Since the bimodal model
doubles the parameter space over which fits must be made, making a comprehensive parameter fit unfeasable, we approximate the statistical uncertainty from the fit to be the same as in the single model fit, about 20\%.

These surface gas masses are compared to the cluster's total mass in the same
volume implied by the strong
gravitational lensing measurements.  The surface total mass can be inferred from a model of the cluster mass distribution which predicts the observed gravitational lensing.
We calculate the gravitational mass using the lensing model in Kneib \etal\ 
(1993) for the same volume, centered at the SZ model fit center.  Although the uncertainties on the model parameters are less than ten percent, we make a conservative estimate of 20\% for the uncertainty, to allow for variations of this model.
Comparing this mass, $9.6^{+1.9}_{-1.9} \times 10^{13} h^{-1} M_{\odot}$,
to the single-component gas mass yields a gas mass fraction, $f_g$, of
$(0.056^{+0.017}_{-0.015})h^{-1}$; the two-component model yields a
slightly lower gas mass fraction of $(0.046\pm 0.013)h^{-1}$.   The
cluster's angular diameter distance was calculated assuming \omegam\ =
0.3 and \omegal\ = 0. 
If \omegam\ = 1.0, \omegal\ = 0, the angular diameter distance to A370
will be 6.5\% smaller, as will $f_g$.  If \omegam\ = 0.3 and \omegal\
= 0.7, $f_g$ will be $\sim 9\%$ larger.

\subsection{Comparison of SZ Gas Mass to Hydrostatic, Isothermal Mass Estimates}
\label{subsec:virialmass}
Since strong gravitational lensing in clusters is relatively rare and is 
restricted to the cores of clusters, and weak lensing analyses are published
for only selected clusters, we also consider
the more general means of calculating the cluster gas
fraction, deprojection of the density model and the HSE assumption.  
We assume the gas is in hydrostatic equilibrium, is isothermal, and is
spheroidally symmetric; for simplicity, we assume the symmetry axis is in the plane of the sky.  The surfaces of
constant electron density are then concentric ellipsoids.
 We then compare the gas mass and the HSE mass from the deprojected
model to determine the gas mass fraction.

Specifically, to determine the gas mass, we extract the central
electron density, $n_{e\circ}$, from the deprojected \be-model and
measured electron temperature by performing the integral along the
line of sight in Equation~\ref{eq:DT0} and integrating
Equation~\ref{eq:density}.  We do this for each set of \be, \rc, and
\dt.  These calculations require
an assumption about the geometry of the cluster, since the core radius 
and extent of the cluster along the line of sight are not known.  
  For an oblate ellipsoid, the axis of symmetry is the cluster's minor axis, and the core radius in the line-of-sight direction is equal to the cluster's observed 
major axis; for a prolate ellipsoid, the axis of symmetry is the major axis, 
and the core radius in the line-of-sight direction is equal to the minor axis.
  
The fit parameters are then used to constrain the total mass.  Hydrostatic equilibrium implies

\begin{equation}{{1 \over \rho_{gas}}\nabla p_{gas} = -\nabla \Phi},\label{eq:hydro}\end{equation}
where $\rho_{gas}$ and $p_{gas}$ are the gas density and pressure, respectively.
The cluster's gravitational potential, $\Phi$,
 can be related to the total mass density, $\rho_{grav}$,
 by Poisson's equation, 

\begin{equation}{\nabla ^2 \Phi = 4\pi G \rho_{grav}},\label{eq:Poisson's} \end{equation} where $G$ is the gravitational constant.  We can solve for the density of the cluster's gravitational mass by 
combining Equations~\ref{eq:hydro} and \ref{eq:Poisson's}:

\begin{equation}{\rho_{grav} = -{1 \over {4 \pi G}}\nabla \cdot \left ({1 \over \rho_{gas}} \nabla p_{gas} \right )}. 
\label{eq:rho_grav1} \end{equation}

We relate the pressure, density, and temperature of the gas through the equation of state:

\begin{equation}{p_{gas} = {{\rho_{gas} k T_e} \over {\mu m_p}}},\label{eq:eofstate} \end{equation}
where $k$ is Boltzmann's constant, $\mu$ is the mean molecular weight of the gas, and $m_p$ is the proton's mass.  To calculate $\mu$, we again assume the gas has the solar metallicity of
Anders and Grevesse (1989) and that $\mu$ is constant throughout the gas.  Making the assumption that the gas is isothermal, we write Equation~\ref{eq:rho_grav1} in the form:

\begin{equation}{\rho_{grav} = -{{k T_e} \over {4 \pi G \mu m_p}}\nabla ^2 {\rm ln}\rho_{gas}}.\label{eq:rho_grav2} \end{equation}
Note that the gravitational mass density depends only on the shape of the gas distribution, and 
so is independent of the value of the central gas density and the gas
mass fraction.  Using the derived shape parameters, \be, \rc, and projected axis ratio, $a$,  and the measured gas temperature, we estimate the total mass density
of the cluster.  We again choose the simplest geometries, that of oblate and prolate ellipsoids, and integrate the density within the same volume as we do the gas density.

For the same range of \be, \rc, and \dt\ as we used in the cylindrical geometry
analysis, we calculate
the cluster's ellipsoidal gas mass, HSE mass, and gas mass fraction
for both prolate and oblate geometries, and the \dchi\ from the best fit parameters.   From \dchi\ = 1 range of fit parameters, we 
derive the $68\%$ confidence intervals for the gas mass and the mean gas mass 
fraction calculated within different major axis radii.

We prefer to measure the masses and mass fractions in the largest volume permitted by our method, since the fair sample assumption is best at large radii and the cores
of clusters may be affected significantly by physical processes not
included in our HSE model (cooling flows, galaxy winds, magnetic fields).  
The largest scales on which we make our calculations are determined by
the shortest baselines on which we detect the SZ effect.
We calculate the statistical uncertainties in the $f_g$ measurement due to the shape parameter uncertainties on a number of scales, from 
10$''$ to 150$''$.  There is a broad minimum in uncertainty around
radius 65$''$.

We calculate the gas mass fraction for the oblate and prolate
spheroids at semi-major axis $65''$ ($\sim 330 h^{-1}$kpc).  The gas
mass is the same for both oblate and prolate geometries;  the change
in central density and volume when the geometry changes exactly
compensate.  For the oblate ellipsoid, the gas mass fraction is
(0.064$_{-0.024}^{+0.024})h^{-1}$; for the prolate ellipsoid, the gas
mass fraction is (0.096$_{-0.034}^{+0.036})h^{-1}$.  We also calculate
$f_g$ in a spherical volume of radius $65''$, using the fits and
uncertainties from the spherical fits and find $f_g$ = (0.080$^{+0.04}_{-0.041})h^{-1}$.  The calculated values are compiled in Table 1.

%\placetable{table:results}  

We also calculate the gas mass fraction for both components of the bimodal model, using a gas temperature of 6.6 keV for
each and find that the gas mass fraction for each component is $\sim 0.040h^{-1}$.

The dark matter needed to produce concentrically ellipsoidal isopotential 
surfaces has a significantly more aspherical distribution than the 
potential.  For the observed axis ratio of 0.64, this dark
matter distribution is unphysical beyond a few core radii in the direction
of the unique axis, as it requires the dark matter density to be
negative.  This also implies that the gas mass fraction will vary
spatially.  The difficulty of this model supporting very elliptical
gas density distributions may be
suggesting that the cluster is bimodal, but it is more likely an indictment of the simplified models we have used.  We have considered the class of 
potentials with isopotential surfaces following concentric ellipsoids 
because they adequately describe the data and because their 
deprojections are straightforward.  Such simplified 
models are seriously deficient, however, for the hydrostatic analysis.  
To produce a cluster with an observed axis ratio of less than $1/\sqrt{2}$ 
in this formalism, the cluster mass must be partially comprised of 
dark matter with negative density.  It is improper, then, to measure the dark matter with this method out past a few core radii for a highly elliptical cluster.  

The effect of using this simple ellipsoidal model to calculate $f_g$ is assessed in the following way.  We construct a
model cluster which has a physically motivated mass
structure, predict its observed SZ effect, and then attempt to recover
the simulated cluster's mass using the same observation and analysis
protocol used for the true observations.  We arrange the dark matter
of the simulated cluster in concentrically spheroidal shells with a
\be-model profile.  The dark matter's axis ratio is set to 2.0:3.5.
Isothermal gas at 7 keV is added in hydrostatic equilibrium with the
cluster potential, and a simulated two-dimensional SZ decrement map
constructed, assuming the cluster is at redshift $z$ = 0.37.  The
decrement map is sampled with the \uv\ coverage of a typical
interferometric observation.  Noise typical of a $\sim$ 40 hour
observation is added.  The resulting \uv\ data are fit in the same
manner described in Section~\ref{sec:model}, with the data fit to a
concentrically ellipsoidal \be-model, and the gas mass and gas mass fraction determined with the HSE method.  Two simulations are used, one model cluster is an oblate spheroid and the other a prolate spheroid, both with the symmetry axis in the plane of the sky.  In both cases, the best fit axis ratio of the gas distribution was 0.79.
These gas masses and gas mass fractions are compared to the actual model values.  These are shown in Table 2.  
%\placetable{table:elliptest}

The model appears to be adequate at small radii, but deteriorates
noticeably by a radius of 100$''$.  This suggests it is reasonable to
measure the gas mass fraction with this model near the same radius we
constrain the data well. We restrict our analysis of the real
observations to within radius of $65'' (\sim$ 330$h^{-1}$ kpc), a region within which the approximation is valid.

\subsection{Systematic Uncertainties}
\label{subsec:systematics}

\subsubsection{Comparing the Lensing and HSE Masses}
If the assumptions made in the hydrostatic isothermal analysis are
valid, the HSE predicted mass should equal the cluster's lensing mass.  We integrate
the total mass density in Equation~\ref{eq:rho_grav2} in the same cylindrical
volume in which the lensing mass is calculated.  This comparison is
potentially a means to discriminate between oblate and prolate models
for the cluster, but, as discussed in
Section~\ref{subsec:virialmass}, our model for ellipsoidal clusters breaks down at large radii.
For the spherical model, this mass is calculated for each set of shape 
parameters and the 68\% confidence limits are derived.  
Using an electron temperature of $6.6^{+1.1}_{-0.9}$ keV, the
spherical geometry HSE mass integrated in the 40$''$ radius cylindrical volume is $1.47^{+0.49}_{-0.38} \times 10^{14} M_\odot$, consistent with the lensing mass in the same geometry 
of $1.59^{+0.30}_{-0.30} \times 10^{14} h^{-1} M_{\odot}$.  
The two-component \be-model predicts about $1.25 \times 10^{14}
M_{\odot}$, and so it is also a consistent model.

This agreement does not guarantee that the
assumptions in the HSE analysis are valid, however.  For this reason, we examine 
possible systematic effects. Some comparisons of cluster total masses derived by the HSE method to lensing masses suggest that the HSE method systematically underpredicts 
the cluster's mass (\eg, Miralda-Escud\'e \& Babul 1994, Loeb \& Mao 1994, Wu \& Fang 1997). 
Some of the explanations suggested for the discrepancy include cluster ellipticity, non-thermal pressure support of the gas in the cluster core, multi-phase gas, and temperature gradients in the gas.  In an examination of a large 
sample of clusters, however, Allen (1998) suggests such discrepancies
can often be resolved by taking account of cooling flows when
analyzing x-ray data, and ensuring that the lensing and x-ray masses
probe the same line of sight.  The mass discrepancy does remain for
clusters in the Allen sample with small cooling flows or none, perhaps suggesting these clusters have undergone recent dynamical activity which have disrupted any pre-existing cooling flow. Such activity might invalidate the HSE assumption.  

We have been careful to ensure that the strong lensing and SZ models probe the same lines of sight. There is no evidence for a cooling flow in Abell 370, although there may be other concerns about using the measured emission-weighted gas temperature.

\subsubsection{Contamination of Emission-Weighted Temperature}
If the measured average temperature is in error, \ie, due to
contamination from a nearby AGN, the mass measurements will also be in error.
The SZ gas mass is inversely proportional to the assumed temperature
and the HSE mass is directly proportional to the temperature.  The gas
fraction from the HSE method is quite sensitive to temperature, $f_g
\propto 1/T_e^2$.  

\subsubsection{Polytropic Temperature Gradient}

An unresolved temperature gradient in the gas may systematically
affect the gas and HSE masses.  If such a gradient is present, 
the true temperature in the central region may be higher than the 
emission-weighted temperature we use, and the fitted shape
parameters from the isothermal SZ analysis may no 
longer accurately describe the density distribution.  

However, if the temperature of the intracluster medium declines
slowly, and does not change appreciably over the angular scales to
which we are sensitive, the interferometric measurement of the gas
mass fraction at these angular scales will in fact not be strongly
affected.  

Currently, there are no strong observational constraints on 
temperature structure in moderately distant and distant clusters, as
there have been no suitable telescope facilities for the task.
However, an attempt to quantify temperature structure observed in 
nearby clusters is presented in Markevitch (1996).  In this work, a 
slow decline in temperature with radius is observed, with the
temperature falling to half its central value at 6-10 core radii.
This structure may be approximately described by a gas with a
polytropic index of $\gamma$ = 1.2.  
(For discussion of polytropic indices, see Sarazin (1988) and the
references within.)  If there temperature structure in Abell 370 and
it is of the moderate variety presented in Markevitch (1996), it 
will not be a strong source of systematic uncertainty.

Abell 370 has a long observation scheduled with the
Chandra observatory.  Chandra has the necessary spatial
and spectral resolution to remove the effects of contaminating sources
in the field and with a long observation, could measure the spatial
variation of the cluster temperature, should it exist.

\subsubsection{Inclination Angle}
The assumption that the cluster is biaxially symmetric with symmetry
axis in the plane of the sky is certainly a simplification.  Here, we
estimate the effect on the gas mass fraction of an inclined symmetry axis.  Analytical
relationships between the inclination angle and the apparent shape of
an biaxial ellipsoid are derived in Fabricant, Rybicki, \& Gorenstein (1984).  We derive the cluster gas mass and HSE mass for a cluster which reproduces
the observed SZ map, but has an inclined symmetry axis.  A biaxial
cluster, when inclined, will retain a \be-model distribution with the
same value of \be, but 
the central density and intrinsic axis ratio will change.   The inclination angle $i$ is measured between the symmetry axis and
the line of sight; $i=90^\circ$ for the analysis of Section~\ref{subsec:virialmass}.  

As inclination
angle increases, so does the intrinsic axis ratio.\footnote{The relationship between intrinsic axis ratio, $a_i$, and the inclination angle, $i$, for an 
oblate ellipsoid is $a_i = a {[1 - a^{-2} \cos^2(i)]^{1/2} \over
\sin(i)}$, where $a$ is the observed axis ratio.  For a prolate
ellipsoid, $a_i = a {\sin i \over [1 - a^{2} \cos^2(i)]^{1/2}}$.}  In
order to preserve the observed SZ effect, The
value of the central density must vary inversely with the axis ratio.
The gas mass, then, 
remains the same for all inclination angles.  This 
is expected, as the gas mass is proportional to the total SZ flux,
independent of its spatial arrangement.  We evaluate the gas and HSE
masses  as a function of $i$, using the best fit parameters from A370.
The gas mass fraction changes, but it changes significantly over a relatively
small range of allowed inclination angles (see Figure 5). 

%\placefigure{fig5}

\section{Discussion}
\label{sec:discussion}
\subsection{Gas Mass Fractions at $r_{500}$}

    To compare the $f_g$  we have measured within a fixed angular
radius to $f_g$ measurements in clusters with different sizes and redshift, we extrapolate our measured $f_g$ to a fiducial radius.  The ICM in nearby clusters has been observed to be distributed more uniformly than
the dark matter (\eg, David \etal\ 1995), which is to be expected if
energy has been added to the intracluster medium before collapse or from
galactic winds.  If this is generally true, the gas mass fraction
measured depends on the radius within which the measurement is made. 
As suggested in Evrard (1997) and Metzler, Evrard, \& Navarro (1998), 
we choose this radius to be that within which the average density of the cluster is 500 times the critical density, $\rho_c = {{3 H^2} \over {8 \pi G}}$.  These numerical simulations suggest that within
this radius, $r_{500}$, the cluster's baryon fraction should closely reflect the universal baryon fraction, if the current physical
models of hierarchical structure formation are correct.  We use the
analytical expression of Evrard (1997) which describes the expected
variation of $f_g$ with overdensity.  This variation is found to be consistent with the $f_g$ variation reported in the David \etal\ (1995) sample.  
\begin{equation}{f_{g}(r_{500}(T_e)) = f_{g}(r_X)\left ({r_{500}(T_e) \over r_X} \right )^{\eta}},\label{eq:fextrap} \end{equation}
where $\eta$ = 0.17, $f_{g}(r_{500}(T_e))$ is the gas mass fraction at $r_{500}$,
and $r_X$ is the radius within which the gas mass fraction is measured.  We
modify Evrard's expression for $r_{500}$, derived for low redshift
clusters, to include the change in the value of $\rho_c$ with
redshift; $\rho_c(z) = \rho_c(z=0)(H/H_\circ)^2$, where 
$H^2=H_\circ^2[(1+z)^3\Omega_M+(1+z)^2(1-\Omega_M-\Omega_\Lambda) + \Omega_\Lambda]$
\begin{equation}{r_{500}(T_e) = (1.24 \pm 0.09) \left ( {T_e \over 10\ {\rm keV} (H/H_\circ)^2} \right ) ^{1/2} h^{-1} {\rm Mpc}}.\label{eq:r500} \end{equation}

The gas mass fraction values at $r_{500}$, estimated from those measured at 65$''$, are summarized in Table 1.

This experiment best measures $f_g$ at a given angular scale, 
which corresponds in Abell 370 to an overdensity of $\sim 5000\rho_c$.  
This is not the optimal radius at which to compare with numerical simulations,
since resolution is limited in the cores of the clusters, and
the gas in the core may also be sensitive to additional physics not yet included in the models, \eg, magnetic fields and cooling.   For these reasons,
the corrections should be taken with some caution.

\subsection{Constraints on \omegam\ from $f_g$}
Under the fair sample hypothesis, A370's gas mass fraction within $r_{500}$, a lower limit to the cluster baryon fraction, should reflect the universal baryon fraction:
\begin{equation} f_g \le f_B = {\Omega_B \over \Omega_M},\end{equation}
where $f_B$ is the cluster's baryon fraction, \omegam\ is the ratio of the total mass density to the
critical mass density, and  $\Omega_B$ is the ratio of baryon mass
density in the universe to the critical mass density.  The cluster gas
mass fraction measurements can then be used within the Big Bang Nucleosynthesis (BBN) paradigm to constrain \omegam:
\begin{equation}\Omega_M \le \Omega_B/f_g.\end{equation}

The value of $\Omega_B$ is constrained by BBN calculations and the
measurements of light element abundances.  The relative 
abundance of deuterium and hydrogen provides a particularly strong constraint on the baryonic matter density.  A firm upper limit to $\Omega_B$ is set by the presence of deuterium in the local interstellar medium.  This constrains the value of $\Omega_B$ to be less than $0.031 h^{-2}$ (Linsky \etal\ 1995).  Measurements of the D/H ratio in metal-poor Lyman-$\alpha$ absorption line systems in high-redshift quasars put a tighter constraint on the baryonic mass density.  Such measurements made by Burles \& Tytler (1998) predict a value of $\Omega_B = (0.019\pm 0.002)h^{-2}$ at 95\% confidence.
 
   The  gas mass fractions  measured  for  A370  from both  lensing and HSE
methods range from $5-13\%h^{-1}$.  We consider the simplest $f_g$
measurement, that in the spherical model, and compare this gas mass
fraction at $r_{500}$ to  the Burles \& Tytler (1998) value for
$\Omega_B$.  This gives an upper limit to the matter density
parameter, $\Omega_M \le 0.19^{+0.10}_{-0.10}h^{-1}$, at 68\%
confidence.  However, the bimodal model gives a surface gas mass
fraction at angular radius 40$''$ of 0.048$h^{-1}$, a value which
permits $\Omega_M$ to be as high as 0.40$h^{-1}$ in this scheme.  (We
note again that the dependence of $f_g$ through the angular diameter
distance is weak at this redshift, with a change in $f_g$ of 5-10\% when a wide range
of cosmological parameters is used.)
These values are consistent with the limits on \omegam\ from
observations of supernovae, which are derived from geometrical
arguments, rather than the $\Omega_B/f_g$ ratio.
Depending on the method used to calibrate the sample, for
a spatially flat universe, Garnavich \etal\ (1998) find \omegam\ $<
0.4-0.5$ at 68\% confidence.

\subsection{Conclusion \& Future Work}	

We have measured the Sunyaev-Zel'dovich effect in the galaxy cluster
Abell 370 and present spatially filtered images from these
data.  The optical and x-ray observations of this cluster show a complicated and 
perhaps bimodal mass distribution.  The SZ effect image, however, looks 
smoothly distributed and significantly aspherical. We have fit both one-component and
two-component \be-models to the data and find that the two-component
model does not fit significantly better.  For both models, we calculate the gas mass fraction
for the cluster using measurements of the total cluster mass from both the
gravitational lens model (the ``surface'' gas mass fraction)  and from
the hydrostatic equilibrium assumption.  When
integrated in the same volume, the HSE masses are consistent with the
mass derived from the gravitational lensing model 
for both the one- and two-component models, lending support to the HSE
assumption.  The surface gas mass fraction measurement is made within an
 angular radius of 40$''$ and the HSE gas mass fraction is made
within a radius of 65$''$.  The gas mass fraction near the virial radius
is derived from the gas mass fractions at 65$''$ using a correction factor 
derived from numerical simulations.
For the range of methods and models used, we find gas mass fraction values of
$\sim (5-13)h^{-1}\%$.  

Constraints on the Hubble parameter \Ho\ can in principle be derived
from the SZ and x-ray measurements of a cluster.  The SZ and x-ray observables depend on different moments of the electron density, and so the characteristic length scale of the cluster along the line of sight can be measured and the angular diameter distance of the cluster inferred.  This is useful not only as a distance ladder-independent
measurement of \Ho, but, when compared with other \Ho\ measurements, 
can be used to explore possible systematic effects in the $f_g$
calculation, \eg, to constrain the deprojection of the gas distribution.
However, the quality of the x-ray imaging data in this case and the
apparent disagreement between the SZ and x-ray fitted models do not
permit putting a strong constraint on the Hubble constant.  A long
observation with the Chandra x-ray observatory towards this cluster is
planned, and should help resolve these issues.

The value of the baryonic mass fraction in any one cluster 
will be susceptible to systematic uncertainties which may be difficult
to estimate.  In order to use cluster gas mass fractions as a
cosmological tool, one wants to ameliorate the effect of these errors 
by studying a large sample of clusters.   A SZ effect survey in galaxy
clusters is being carried out by
this group at the BIMA and OVRO observatories, and an analysis of the gas
mass fraction of the sample is in preparation.

	Many thanks are due to the staff at the BIMA and OVRO
observatories for their contributions to this project, especially Rick
Forster, John Lugten, Steve Padin, Dick Plambeck, Steve Scott, and
Dave Woody.  Many thanks to Cheryl Alexander for her work on the
system hardware.  Thanks also to 
Jack Hughes and Doris Neumann for valuable discussions concerning the x-ray 
analysis, to Jean-Paul Kneib concerning his lensing analysis, 
and to Naomi Ota and collaborators for sharing their reduced ASCA data.
This work is supported by NASA LTSA grant NAG5-7986. 
LG, EDR, and SKP gratefully thank the NASA GSRP program for its
support.  Radio astronomy with the OVRO and BIMA millimeter arrays is
supported by NSF grant AST 93-14079 and AST 96-13998, respectively.
The funds for the additional hardware for the SZ experiment were from a NASA CDDF grant, a NSF-YI Award, and the David and Lucile Packard Foundation.

\newpage

\begin{deluxetable}{llll}
\tablewidth{0pt}
\tablenum{1}
\tablecaption{Gas Mass Fractions for Abell 370\label{table:results}}
\tablehead{
\colhead{geometry} 	& \colhead{$f_g$}	 & 
\colhead{$\frac{f_g(\delta_{500})}{f_g(65")}$}	 & \colhead{$f_g(\delta_{500})$}}
\startdata
cylinder (radius=40$''$) & $(0.056^{+0.017}_{-0.015})h^{-1}$ & ...  & ... \\
cylinder (radius=40$''$), bimodal & $(0.046^{+0.013}_{-0.013})h^{-1}$
& ... &... \\ 
oblate ellipsoid (radius=65$''$) & (0.064$_{-0.024}^{+0.026})h^{-1}$ & 1.22 & (0.078$_{-0.029}^{+0.032})h^{-1}$\\ 
prolate ellipsoid (radius=65$''$)& (0.106$_{-0.044}^{+0.048})h^{-1}$ & 1.22 & (0.129$_{-0.054}^{+0.059})h^{-1}$\\ 
sphere (radius=65$''$) & (0.080$^{+0.044}_{-0.041})h^{-1}$& 1.22 & (0.098$^{+0.050}_{-0.054})h^{-1}$\\
\enddata
\tablecomments{The surface $f_g$
(cylindrical geometry) is derived from the elliptical \be-model (with
$a=0.64$) and the two-component model for the SZ effect and the total mass from the strong
gravitational lensing model of Kneib \etal\ (1993).  The ellipsoidal
gas mass fractions are calculated from deprojections of both the elliptical and
spherical \be-model fits to the SZ data and the
isothermal HSE assumption. The major axis indicated in the first
column and symmetry axes are assumed to be in the plane of the sky.
The radius within which $f_g$ is measured is indicated in the first
column; $f_g$ within $r_{500}$ is estimated in Column 4,
using a correction factor (Column 3) from Equation~\ref{eq:fextrap}.}
\end{deluxetable}

\begin{deluxetable}{lcc}
\tablewidth{0pt}
\tablenum{2}
\tablecaption{Comparison of Derived to True Gas Mass and Gas Mass Fraction\label{table:elliptest}}
\tablehead{
\colhead{} & \colhead{Within 65$''$} & \colhead{Within 100$''$}}
\startdata
{Oblate Cluster}, $i=90^\circ$ & & \\
 & {derived value/true value} &{derived value/true value}\\
total mass  & 1.05$^{+0.08}_{-0.08}$& 1.42$^{+0.16}_{-0.43}$\\
gas mass &0.98$^{+0.08}_{-0.08}$ & 0.96$^{+0.14}_{-0.12}$\\
gas mass fraction &0.95$^{+0.19}_{-0.19}$&0.65$^{+0.35}_{-0.06}$ \\
%\tablevspace{0.2in}
\\
{Prolate Cluster}, $i=90^\circ$ & &\\
& {derived value/true value} &{derived value/true value}\\
total mass  & 0.88$^{+0.17}_{-0.17}$ & 1.16$^{+0.27}_{-0.14}$\\
gas mass  & 0.87$^{+0.26}_{-0.07}$ &0.88$^{+0.34}_{-0.14}$\\
gas mass fraction &1.11$^{+0.62}_{-0.25}$  & 0.79$^{+0.87}_{-0.20}$\\
\enddata
\\
\tablecomments{We test the effect of using an ellipsoidally
symmetric \be-model for the gas density distribution.  We simulate SZ observations
of an oblate cluster and a prolate cluster and analyze them with the methods of
Sections ~\ref{sec:model} and ~\ref{subsec:virialmass}.  We compare
the quantities we derive to the true simulated cluster's values.}
\end{deluxetable}

\begin{figure}
\figurenum{1}
\epsscale{0.75}
\includegraphics[angle=270,scale=0.8]{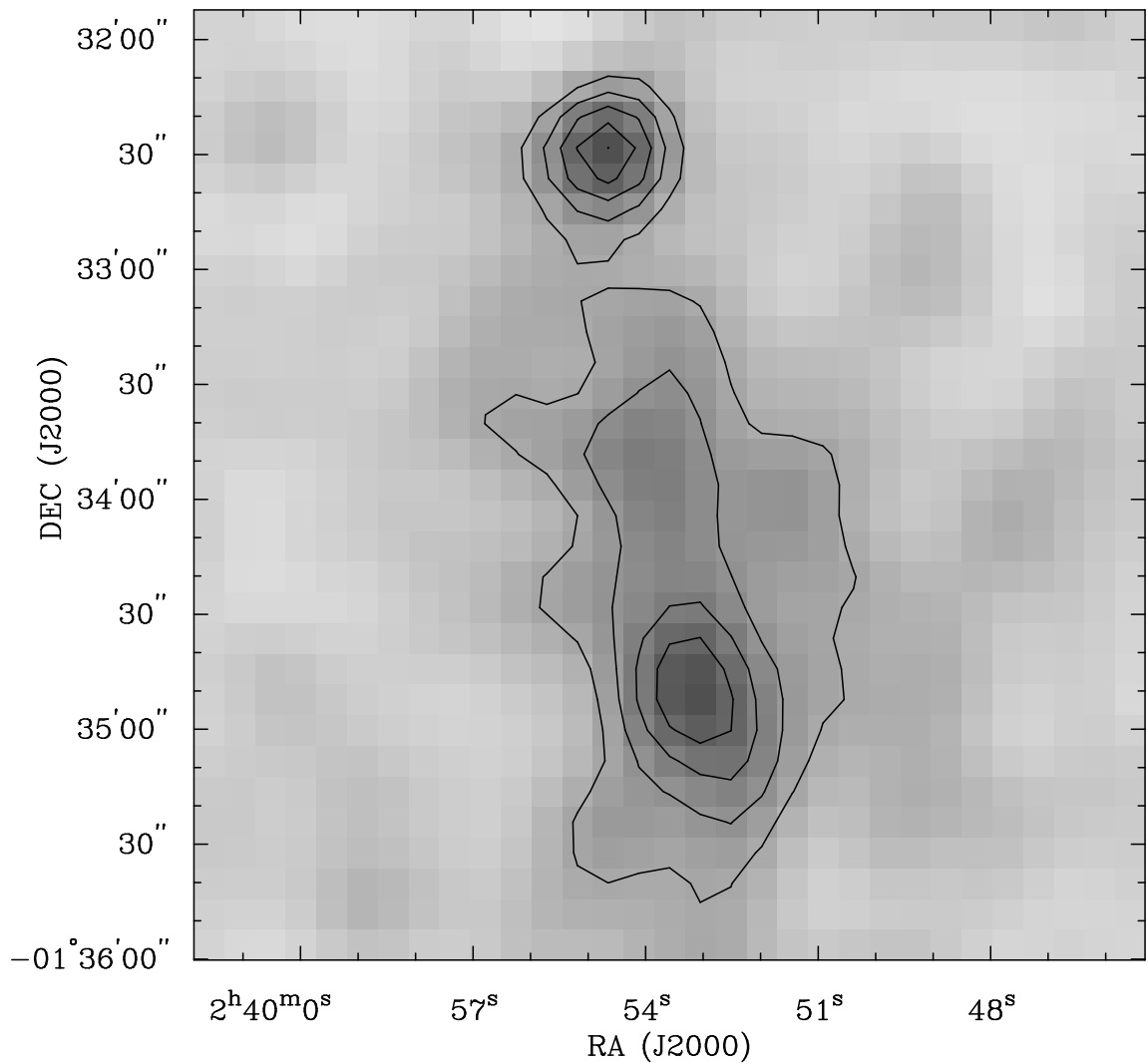}
\caption{X-ray image of the galaxy cluster Abell 370 observed with the
{\em ROSAT} HRI.  The x-ray image was filtered to include only PHA channels 1-7, binned in 8$''$ x 8$''$ pixels, and smoothed with an 10$''$ Gaussian filter.  The
background level is 2.8 x 10$^{-3}$ HRI counts sec$^{-1}$ arcmin $^{-2}$
and the countour levels are  at 5.8, 7.2, 8.7, 9.5, 10.1, and 11.6 x
10$^{-3}$ HRI counts sec$^{-1}$ arcmin $^{-2}$.  The x-ray emission peak is
associated with a pointlike source located $\sim2.5'$ to the north of the
cluster center.\label{fig1}}
\end{figure}

\begin{figure}
\figurenum{2}
\epsscale{0.4}
\includegraphics[angle=270,scale=0.8]{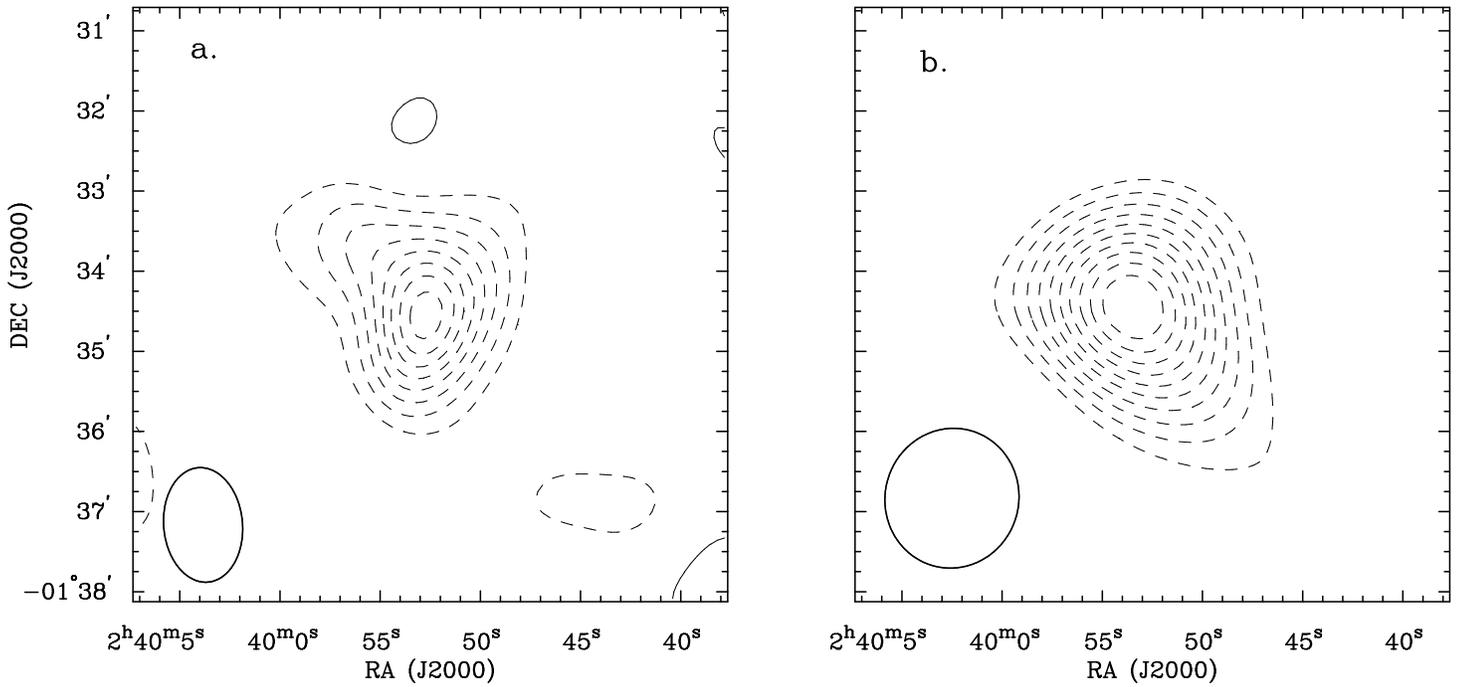}
\caption{a. Image of Abell 370 from OVRO observations.  The restoring
beam, shown in the lower left hand corner, has a Gaussian FWHM of
$59''$ x $86''$.  The RMS noise in the map is 50 $\mu$Jy beam$^{-1}$, or 14.9
$\mu$K, and the integrated flux density of the source is $-1.54\pm0.17$ mJy.  b. Image of A370 from BIMA observations. The restoring beam has FWHM of $91''$ x $95''$.  The RMS noise in
the map is 180 $\mu$Jy beam$^{-1}$, or 31.4 $\mu$K, and the integrated flux
density of the source is $-4.34\pm0.52$ mJy.\label{fig2}}
\end{figure}

\begin{figure}
\figurenum{3}
\epsscale{1.2}
\plottwo{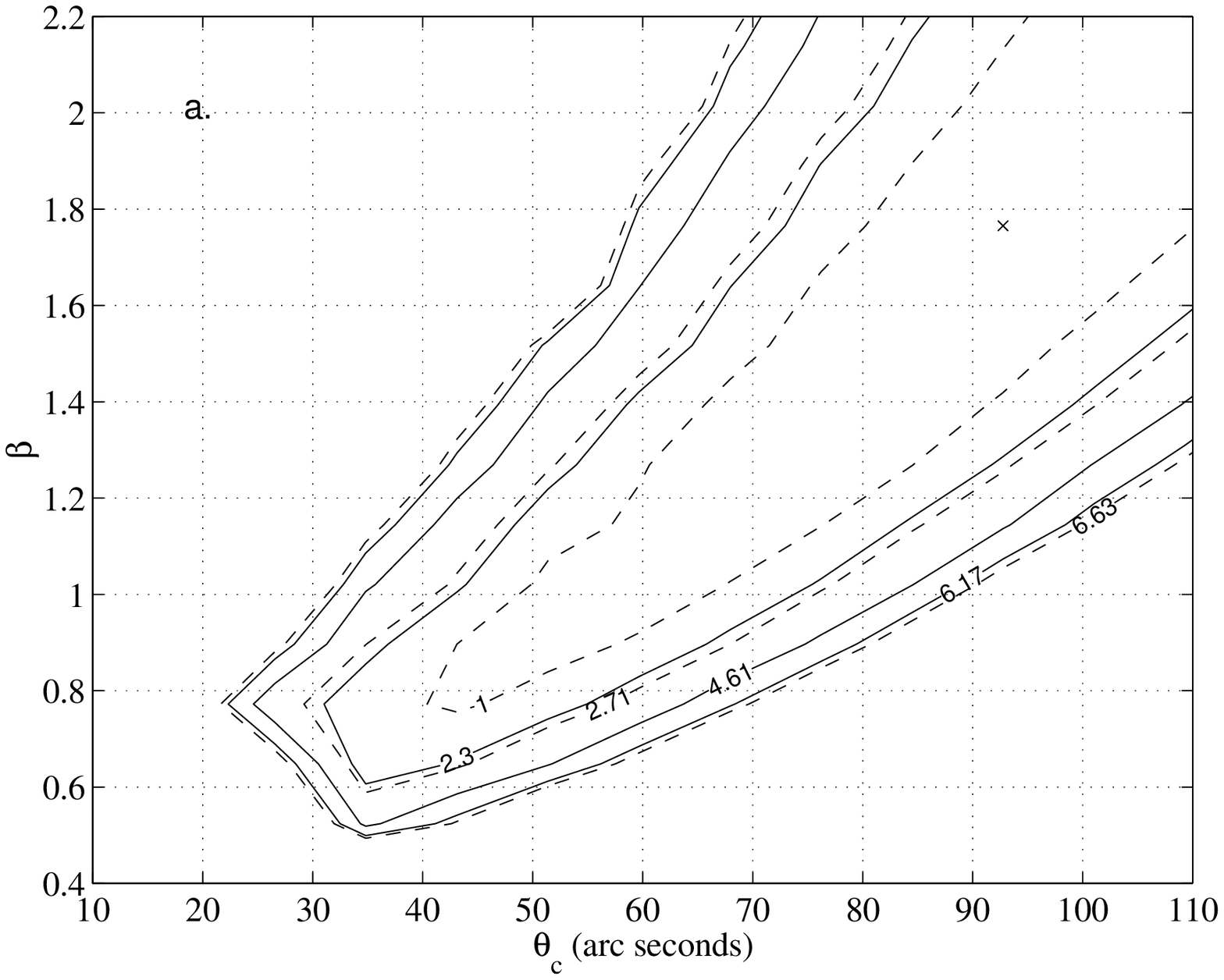}{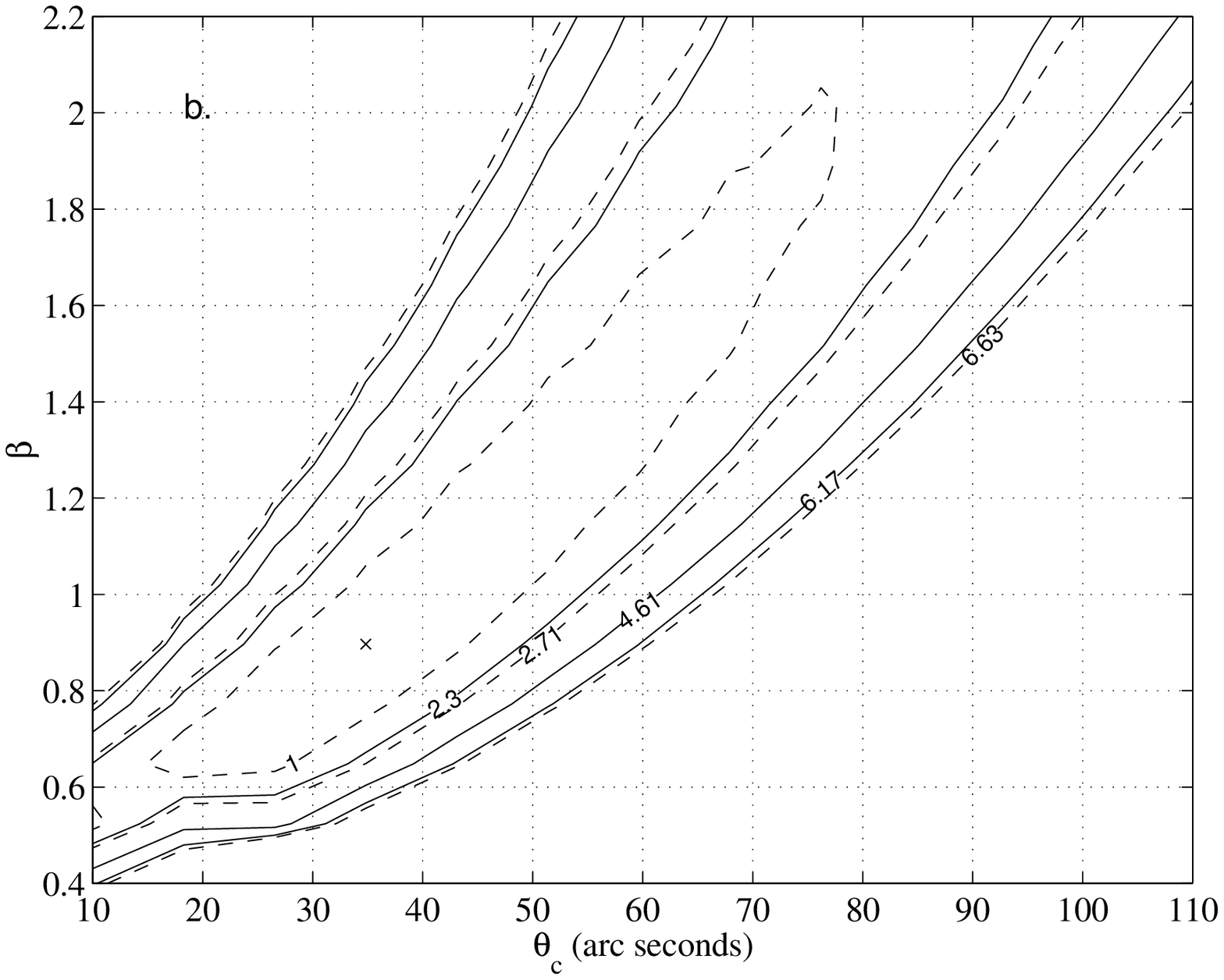}
\caption{Confidence regions from the joint fit to the BIMA and OVRO SZ
data using a. an elliptical \be-model and b. a spherical \be-model.
In both plots, the best fit (\be, \rc) point is indicated with an x
and the full line contours show the \dchi\ = 2.3, 4.61, and 6.17
regions, which indicate 68.3\%, 90.0\%, and 95.4\% confidence, respectively,
for the two-parameter fit.  The dashed lines show \dchi\ $=$1.0,
2.71, and 6.63 regions; the projection of these regions onto the \be\ or \rc\ axis indicate the 68.3\%, 90\% and
99\% confidence interval on the single parameter.  The centroid and point source flux and position are fixed at the
best fit values, and \dt\ is allowed to assume its best fit value at
each (\be, \rc) point.
\label{fig3}}
\end{figure}

\begin{figure}
\figurenum{4}
\epsscale{0.9}
\plotone{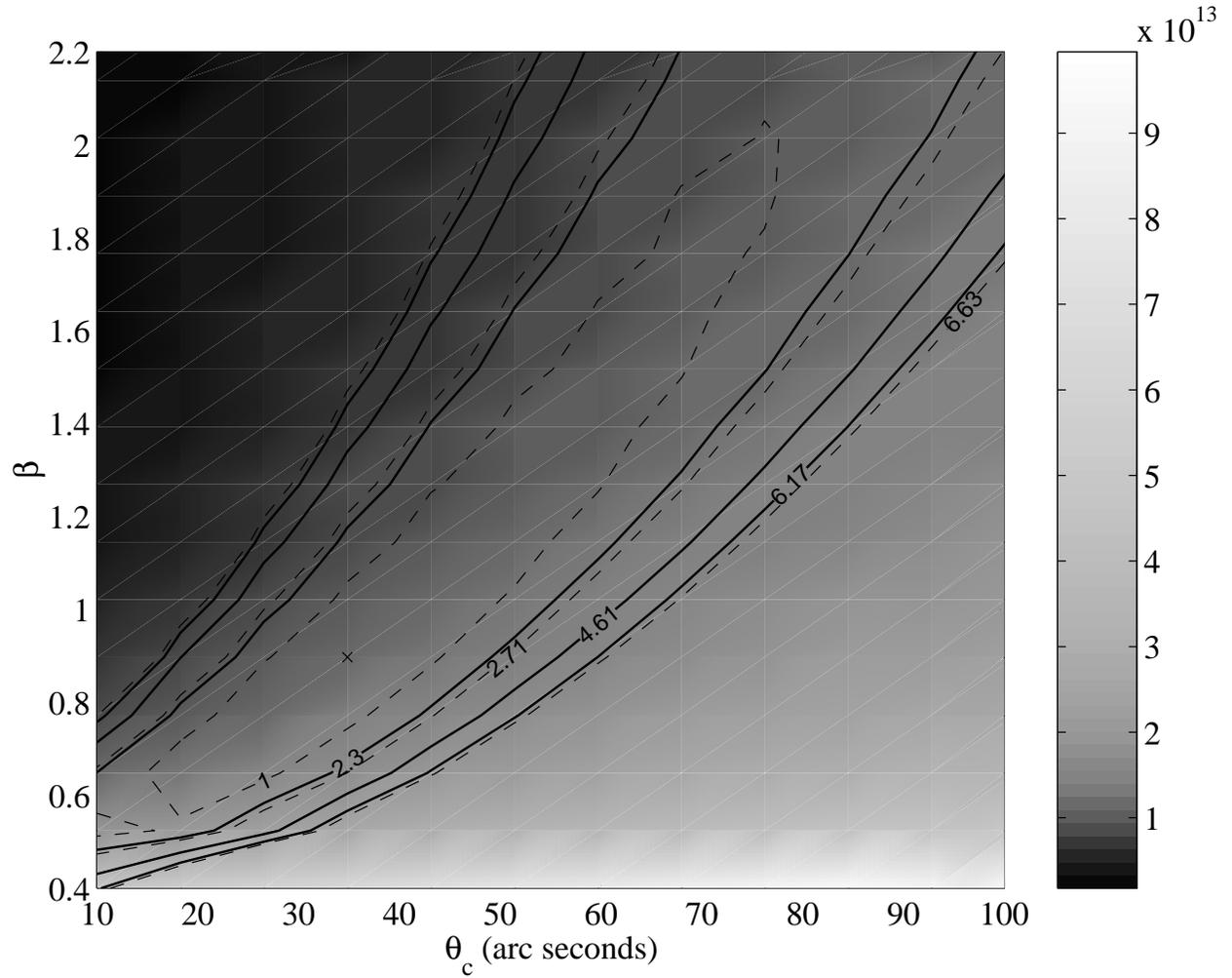}
\caption{Gas Mass for Abell 370.  The gas mass in a cylindrical volume with
cross section of 65$''$ and axis ratio 0.64, in units of $M_{\odot}$ is shown in greyscale.  The two-parameter confidence intervals for $\beta$ and $r_c$ are
overlaid.  At each point, $\Delta T(0)$ assumes its best fit value.
The best fit point is marked with an x.\label{fig4}}
\end{figure}

\begin{figure}[nt]
\figurenum{5}
\epsscale{0.9}
\plotone{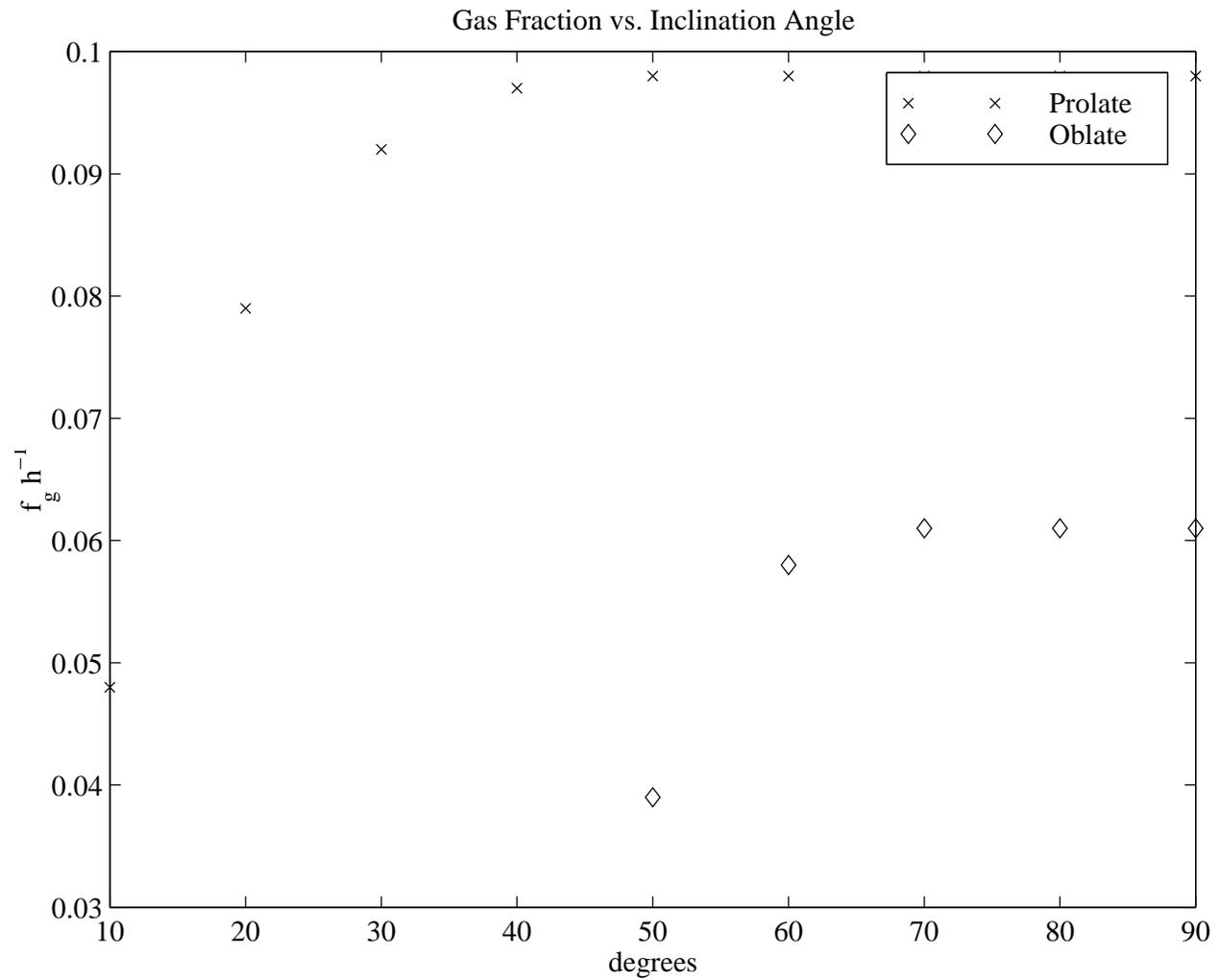}
\caption{The ellipsoidal gas mass fraction, calculated with the
isothermal HSE method, as a function of inclination angle, $i$.  The mass is calculated using the best fit parameters from the elliptical \be-model fit. \label{fig5}}
\end{figure}

\end{document}